\documentclass[aps,prb,twocolumn,superscriptaddress,showpacs,showkeys,amsmath,amssymb]{revtex4}

\usepackage{amsfonts}
\usepackage{amssymb,amsmath}
\usepackage{graphicx}
\usepackage{dcolumn}
\usepackage{bm}
\usepackage{color}

\begin{document}

\title{Localized states on triangular traps
and low-temperature properties of the antiferromagnetic Heisenberg and repulsive Hubbard models}

\author{Mykola Maksymenko}
\affiliation{Institute for Condensed Matter Physics,
          National Academy of Sciences of Ukraine,
          1 Svientsitskii Street, L'viv-11, 79011, Ukraine}

\author{Oleg Derzhko}
\affiliation{Institute for Condensed Matter Physics,
          National Academy of Sciences of Ukraine,
          1 Svientsitskii Street, L'viv-11, 79011, Ukraine}
\affiliation{Institut f\"{u}r theoretische Physik,
          Universit\"{a}t Magdeburg,
          P.O. Box 4120, D-39016 Magdeburg, Germany}

\author{Johannes Richter}
\affiliation{Institut f\"{u}r theoretische Physik,
          Universit\"{a}t Magdeburg,
          P.O. Box 4120, D-39016 Magdeburg, Germany}

\date{\today}



\begin{abstract}
We consider the antiferromagnetic Heisenberg and the repulsive Hubbard model
on two $N$-site one-dimensional lattices, 
which support dispersionless one-particle states 
corresponding to localized states on triangular trapping cells.
We calculate the degeneracy of the ground states
in the subspaces with $n\le n_{\max}$, $n_{\max}\propto N$ magnons or electrons
as well as the contribution of these states 
(independent localized states)
to thermodynamic quantities.
Moreover,
we discuss another class of low-lying eigenstates 
(so-called interacting localized states) 
and calculate their contribution to the partition function.  
We also discuss the effect of extra interactions,
which lift the degeneracy present due to the chirality of the localized states on triangles.
The localized states set an extra low-energy scale in the system 
and lead to a nonzero residual ground-state entropy and to one (or more) additional low-temperature peak(s) in the specific heat.
Low-energy degrees of freedom in the presence of perturbations removing degeneracy owing to the chirality
can be described in terms of an effective (pseudo)spin-1/2 transverse $XX$ chain.
\end{abstract}

\maketitle

\section{Introduction}
\label{sec1}
\setcounter{equation}{0}

The thermodynamics of strongly correlated lattice models is generally unknown.
Although 
analytical 
(like conventional Green's function technique, dynamical mean-field theory etc)
and 
numerical 
(like series expansions, quantum Monte Carlo algorithms, density-matrix renormalization group algorithms etc)
methods being applied appropriately in particular cases 
may yield desired thermodynamic characteristics with required accuracy,
seeking for new approaches permanently attracts much attention of theoreticians.
One interesting idea for calculating thermodynamic quantities for strongly correlated systems,
which is related to the concept of localized one-particle states,\cite{mielke,lm}
has been suggested recently 
for some spin\cite{localized_magnons_a,localized_magnons_b} and electron\cite{localized_electrons} models.
The localized nature of one-particle states for certain classes of lattices
allows to construct exactly the relevant many-particle states 
and to estimate their contribution to thermodynamics using classical lattice-gas models
which are much easier to investigate than the initial quantum many-body models.

In previous investigations of localized states 
performed for the antiferromagnetic Heisenberg and the repulsive Hubbard model 
on highly frustrated lattices\cite{mielke,lm,localized_magnons_a,localized_magnons_b,localized_electrons} 
mainly bipartite trapping cells
(e.g., a single bond or equilateral even polygons)
were considered. 
These cells have a nondegenerate ground state for the one-particle problem.
Here we extend the discussion of localized states to highly frustrated lattices 
with non-bipartite {\it{triangular}} trapping cells.
A non-bipartite cell may have a degenerate ground state for the one-particle problem 
that may lead to new effects.
For example, 
an equilateral triangle has a two-fold degenerate one-particle ground state, 
which can be related to the chirality degrees of freedom associated to a triangle, 
see, e.g., Ref.~\onlinecite{chirality-parameter}. 
To be specific, 
we consider 
(i) a one-dimensional (1D) lattice which consists of corner sharing ``double-tetrahedra'' 
(the double-tetrahedra chain) 
and 
(ii) a frustrated (cylindrical) three-leg ladder having a triangular arrangement of rungs  
(the frustrated triangular tube), 
cf. Fig.~\ref{fig01}.
Both lattice geometries have been considered in the literature, 
see, 
e.g., 
Refs.~\onlinecite{mambrini,rojas,batista,antonosyan} for the double-tetrahedra chain 
and
Refs.~\onlinecite{subrahmanyam,andreas,luescher,tube_schnack,FLP:PRB06,nedko,sakai,penc} for the triangular tube.
Note that triangular-tube geometry is realized 
for the copper spins in [(CuCl$_2$tachH)$_3$Cl]Cl$_2$.\cite{tube_schnack,nedko}

In what follows 
we consider two concrete models of strongly correlated systems on these lattices, 
namely the spin-1/2 Heisenberg model and the Hubbard model,   
and discuss the consequences of the localized-magnon or localized-electron states 
in combination with the additional chirality degrees of freedom.
We mention that some similar ideas have been elaborated recently 
for the Hubbard model on decorated lattices,\cite{batista} 
for the coupled tetrahedral Heisenberg chain\cite{mambrini,rojas}
as well as for the frustrated Heisenberg spin tube.\cite{andreas}
However, in these references 
the concept of localized states has not been used to discuss low-temperature thermodynamics
for thermodynamically large systems.  
Furthermore note that some preliminary results of our study were announced in a conference paper.\cite{submitted}

The paper is organized as follows.
In Sec. \ref{sec2} we discuss the one-particle spectra of the spin and electron models.
In Sec. \ref{sec3} we briefly illustrate the construction of independent localized many-particle states 
and calculate the contribution of these  states to thermodynamic quantities.
Then, 
in Sec. \ref{sec4}, we illustrate how we can go beyond the independent localized states 
taking into account additional low-energy excitations.
Finally,  in Sec. \ref{sec5} we consider symmetry-breaking interactions 
which lift the degeneracy related to the chirality.
We end up with a summary of our findings in Sec. \ref{sec6}.

\section{Heisenberg and Hubbard models on one-dimensional lattices with triangular traps}
\label{sec2}
\setcounter{equation}{0}

\begin{figure}
\begin{center}
\includegraphics[clip=on,width=77mm,angle=0]{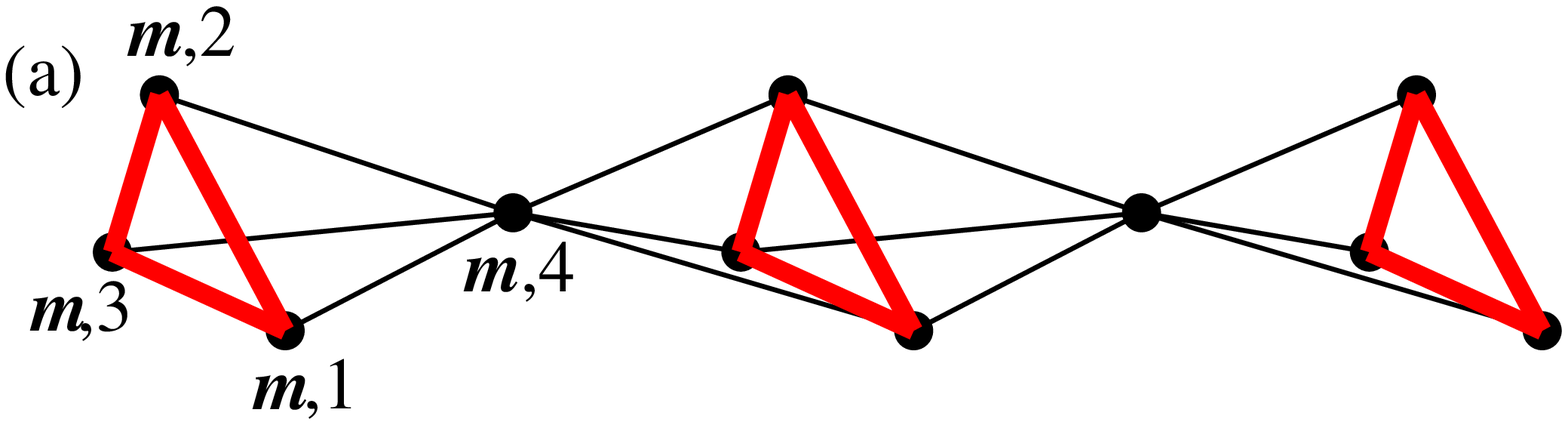}\\
\includegraphics[clip=on,width=77mm,angle=0]{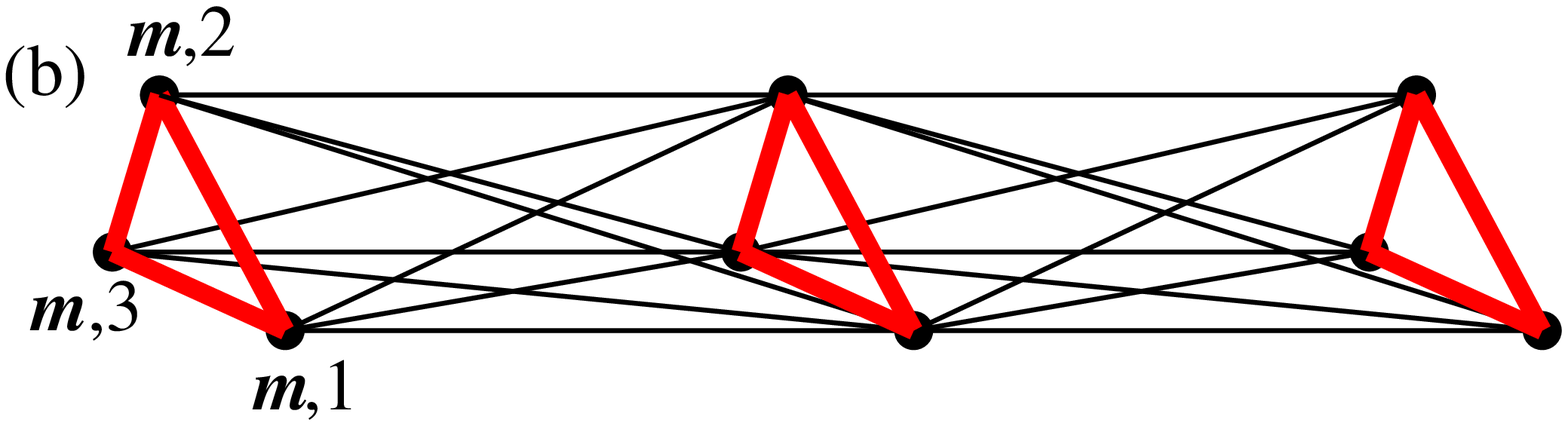}
\caption
{(Color online)
Two 1D frustrated lattices with triangular trapping cells:
(a) the double-tetrahedra chain 
and 
(b) the frustrated (cylindrical) three-leg ladder (or frustrated triangular tube).
The exchange or hopping integrals along the equilateral triangles are $J_2>0$ or $t_2>0$ (bold bonds)
whereas all other exchange or hopping integrals are $J_1>0$ or $t_1>0$ (thin bonds).}
\label{fig01}
\end{center}
\end{figure}

In our study we consider the two 1D lattices shown in Fig.~\ref{fig01}.
The double-tetrahedra chain [panel (a)] may be viewed as a generalization of the diamond chain
(the vertical bond in the diamond chain is replaced by the equilateral triangle).
The frustrated triangular tube [panel (b)] may be viewed as a generalization of the frustrated two-leg ladder
(again the vertical bond is replaced by the equilateral triangle).
The essential geometrical element of the considered lattices are these equilateral triangles
(which act as trapping cells, see below) 
together with the surrounding (connecting) bonds attached to the sites of these equilateral triangles.
In order that the connecting bonds should prevent the escape of the localized magnon (electron) from the triangular trap, 
each bond of the trapping cell together with two of the connecting bonds attached to this trapping-cell bond  
must form an isosceles triangle, 
i.e., the two connecting bonds must be equal to each other.
As a result, 
the considered lattices, owing to destructive quantum interference, support localized one-particle states.
We note here that the lattices with triangular trapping cells may be constructed in higher dimensions too, 
see Refs.~\onlinecite{batista} and \onlinecite{loh}.

On these 1D lattices we consider 
the spin-1/2 Heisenberg antiferromagnet with the Hamiltonian
\begin{eqnarray}
\label{2.01}
H_{\rm{s}}=\sum_{(ij)}J_{ij}{\bf{s}}_i \cdot{\bf{s}}_j-hS^z,
\;\;\;
S^z=\sum_is_i^z
\end{eqnarray}
and the repulsive Hubbard model
\begin{eqnarray}
\label{2.02}
H_{\rm{e}}=\sum_{\sigma=\uparrow,\downarrow}H_{0,\sigma}+U\sum_{i}n_{i,\uparrow}n_{i,\downarrow},
\;\;\;
U>0,
\nonumber\\
H_{0,\sigma}=\sum_{(ij)}t_{ij}\left(c_{i,\sigma}^{\dagger}c_{j,\sigma}+c_{j,\sigma}^{\dagger}c_{i,\sigma}\right)
+\mu\sum_{i}n_{i,\sigma}.
\end{eqnarray}
We use standard notations in Eqs.~(\ref{2.01}) and (\ref{2.02}) and imply periodic boundary conditions.
The exchange or hopping integrals acquire two values:
$J_2>0$ or $t_2>0$ along the equilateral triangles (bold bonds in Fig.~\ref{fig01})
and
$J_1>0$ or $t_1>0$ along all other bonds (thin bonds in Fig.~\ref{fig01}).
It is convenient to label the lattice sites by a pair of indeces,
where the first number enumerates the cells
($m=1,\ldots,{\cal{N}}$,
${\cal{N}}=N/4$ for the double-tetrahedra chain,
${\cal{N}}=N/3$ for the frustrated triangular tube,
$N$ is the number of sites)
and the second one enumerates the position of the site within the cell,
see Fig.~\ref{fig01}.

The one-particle (one-magnon or one-electron) energy spectra for both models 
(with $h=0$ or $\mu=0$) 
can easily be calculated yielding
\begin{eqnarray}
\varepsilon_{1,2}(\kappa)=-\frac{3}{2}J_2-J_1=-\varepsilon,
\nonumber\\
\varepsilon_{3,4}(\kappa)=-2J_1\mp J_1\sqrt{1+\frac{3}{2}(1+\cos\kappa)}
\label{2.03}
\end{eqnarray}
(double-tetrahedra chain)
and
\begin{eqnarray}
\varepsilon_{1,2}(\kappa)=-\frac{3}{2}J_2-3J_1=-\varepsilon,
\nonumber\\
\varepsilon_{3}(\kappa)=-3J_1+3J_1\cos\kappa
\label{2.04}
\end{eqnarray}
(frustrated triangular tube)
for the spin model
and
\begin{eqnarray}
\varepsilon_{1,2}(\kappa)=-t_2=-\varepsilon,
\nonumber\\
\varepsilon_{3,4}(\kappa)=t_2\mp \sqrt{t_2^2+6t_1^2(1+\cos\kappa)}
\label{2.05}
\end{eqnarray}
(double-tetrahedra chain)
and
\begin{eqnarray}
\varepsilon_{1,2}(\kappa)=-t_2=-\varepsilon,
\nonumber\\
\varepsilon_{3}(\kappa)=2t_2+6t_1\cos\kappa
\label{2.06}
\end{eqnarray}
(frustrated triangular tube)
for the electron model.
The flat (dispersionless) bands 
$\varepsilon_{1,2}(\kappa)=-\varepsilon$ 
allow to construct such wave packets of Bloch states 
which are localized on the triangles.
These localized one-particle states read
\begin{eqnarray}
\vert +\rangle_m
=\frac{1}{\sqrt{3}}
\left(s_{m,1}^-+\omega s_{m,2}^-+\omega^2 s_{m,3}^-\right)\vert{\rm{FM}}\rangle,
\nonumber\\
\vert -\rangle_m
=\frac{1}{\sqrt{3}}
\left(s_{m,1}^-+\omega^2 s_{m,2}^-+\omega s_{m,3}^-\right)\vert{\rm{FM}}\rangle,
\label{2.07}
\end{eqnarray}
where $\vert{\rm{FM}}\rangle$ denotes the ferromagnetic background for the spin model 
and
\begin{eqnarray}
\vert +\rangle_m
=\frac{1}{\sqrt{3}}
\left(c_{m,1}^{\dagger}+\omega c_{m,2}^{\dagger}+\omega^2 c_{m,3}^{\dagger}\right)\vert{\rm{0}}\rangle,
\nonumber\\
\vert -\rangle_m
=\frac{1}{\sqrt{3}}
\left(c_{m,1}^{\dagger}+\omega^2 c_{m,2}^{\dagger}+\omega c_{m,3}^{\dagger}\right)\vert{\rm{0}}\rangle,
\label{2.08}
\end{eqnarray}
where $\vert{\rm{0}}\rangle$ denotes the vacuum state for the electron model
and the spin index $\sigma$ is omitted as irrelevant for the one-electron problem.
Here $\omega=e^{2\pi i/3}$.

The two-fold degeneracy of the flat bands corresponds to two possible values of the chirality of the triangle.
For the spin model,
after introducing the chirality operator for a triangle\cite{chirality-parameter}
\begin{eqnarray}
\label{2.09}
\chi_m=\frac{4}{\sqrt{3}}\left({\bf{s}}_{m,1}\cdot\left[{\bf{s}}_{m,2}\times{\bf{s}}_{m,3}\right]\right)
\nonumber\\
=\frac{2i}{\sqrt{3}}
\left[\left(s_{m,1}^+s_{m,2}^--s_{m,1}^-s_{m,2}^+\right)s_{m,3}^z
\right.
\nonumber\\
\left.
+\left(s_{m,2}^+s_{m,3}^--s_{m,2}^-s_{m,3}^+\right)s_{m,1}^z
\right.
\nonumber\\
\left.
+\left(s_{m,3}^+s_{m,1}^--s_{m,3}^-s_{m,1}^+\right)s_{m,2}^z\right],
\end{eqnarray}
we find
$\chi_m \vert\pm\rangle_m = \pm \vert\pm\rangle_m$.
We notice that the $s^z$ operators in Eq.~(\ref{2.09}) yield simply 1/2
after acting of $\chi_m$ on the states $\vert \pm\rangle_m$ (\ref{2.07}).
Therefore we may choose a simpler form of the chirality operator
omitting the operators $s^z$ and the factor 2 in the last expression in Eq.~(\ref{2.09}),
see, e.g., Ref.~\onlinecite{chirality-parameter}.
Similarly, for the electron models
\begin{eqnarray}
\label{2.10}
\chi_m
=-\frac{i}{\sqrt{3}}
\left(c_{m,1}^{\dagger}c_{m,2}+c_{m,1}c_{m,2}^{\dagger}
\right.
\nonumber\\
\left.
+c_{m,2}^{\dagger}c_{m,3}+c_{m,2}c_{m,3}^{\dagger}
+c_{m,3}^{\dagger}c_{m,1}+c_{m,3}c_{m,1}^{\dagger}\right)
\end{eqnarray}
and again $\chi_m \vert\pm\rangle_m = \pm \vert\pm\rangle_m$.
In both spin and electron cases the chirality operator $\chi_m$ can be written in the form
\begin{eqnarray}
\label{2.11}
\chi_m=\vert +\rangle_m\langle +\vert_m - \vert -\rangle_m\langle -\vert_m,
\end{eqnarray}
see Eqs.~(\ref{2.07}) and (\ref{2.08}).

From the above equations for the spectra it is obvious 
that the two-fold degenerate flat band becomes the lowest one 
if 
$J_2> 2J_1$ for the spin model
or 
$t_2>2t_1$ for the electron model. 
In what follows we assume that these  ratios are fulfilled.

\section{The contribution of independent localized states to thermodynamic quantities}
\label{sec3}
\setcounter{equation}{0}

The spin Hamiltonian (\ref{2.01}) commutes with $S^z$, 
i.e., the number of magnons $n=N/2-S^z$ is a good quantum number.
Similarly,
the electron Hamiltonian (\ref{2.02}) commutes with the operator of the number of electrons.
Therefore, we may consider the subspaces with different numbers of magnons or electrons separately.
Moreover,
we may assume at first $h=0$ or $\mu=0$ and add trivial contributions of these terms to the partition function later.

We start with the construction of localized many-particle eigenstates 
in the subspaces with $n\le n_{\max}\propto N $ magnons or electrons 
based on the localized one-particle states.
These localized many-particle states are obtained by occupying the
triangular traps with localized particles.
For the occupation of the traps certain rules have to be fulfilled, 
cf. 
Ref.~\onlinecite{localized_magnons_b} for spin systems 
and
Ref.~\onlinecite{localized_electrons} for electron systems. 
For the spin system on the frustrated-tube lattice localized magnons cannot occupy neighboring triangular traps 
whereas for the double-tetrahedra spin chain the occupation of neighboring triangular traps is allowed. 
Hence, according to Ref.~\onlinecite{localized_magnons_b}, 
the frustrated triangular spin tube belongs to the hard-dimer class 
and 
the double-tetrahedra spin chain belongs to the hard-monomer class.
For both electron models localized electrons may occupy neighboring traps.
Moreover, for the electron system it is possible that two electrons forming a spin-1 triplet 
(e.g., two spin-up electrons) 
but having different chiralities occupy the same triangular trap. 
Note that the different trap occupation rules lead finally to the different relations 
between the maximum number of localized magnons (electrons) $n_{\max}$ and the number of cells $\cal N$ given below.

It is helpful to bear in mind a simple picture 
visualizing this construction of the many-particle states.\cite{localized_magnons_a,localized_magnons_b,localized_electrons}
Namely, 
the construction of the many-particle states may be associated with a filling of an auxiliary lattice 
(a simple chain of ${\cal{N}}$ sites in all cases considered here) 
by hard-core objects 
(hard monomers or hard dimers) 
of two colors corresponding to two values of the chirality.
Moreover, 
for the electron systems we have to take into account in addition the electron spin and the Pauli principle.
Thus, 
the maximum filling with magnons is $n_{\max}={\cal{N}}$ (double-tetrahedra Heisenberg chain) 
and 
$n_{\max}={\cal{N}}/2$ (frustrated Heisenberg triangular tube),
whereas for the Hubbard model the maximum filling with electrons is  $n_{\max}=2{\cal{N}}$
for both lattices.

According to these rules the localized many-particle states are product states of localized one-particle states
with the energy $E_{\rm{FM}}-n\varepsilon$ 
($E_{\rm{FM}}$ is the energy of the ferromagnetic state)
for the spin models
or 
with the energy $-n\varepsilon$ for the electron models, 
where $\varepsilon$ is given in Eqs.~(\ref{2.03}) -- (\ref{2.06}).
Importantly, 
localized electron eigenstates constructed in this way do not feel the Hubbard interaction $U$.
Furthermore, 
the localized many-particle states are the only ground states 
in the corresponding subspaces with up to $n_{\max}$ magnons or electrons
if $J_2>2J_1$ (spin models) or $t_2>2t_1$ (electron models).
We have checked this analyzing full diagonalization data for several finite spin and electron systems.
Obviously, 
there is a large manifold of degenerate localized many-particle ground states in an $n$-particle subspace.
We will denote this ground-state degeneracy in what follows as $g_{\cal{N}}(n)$.

For the spin models the contribution of these localized eigenstates to the partition function is given by
\begin{eqnarray}
\label{3.01}
Z_{{\rm{lm}}}(T,h,N)=\sum_{n=0}^{n_{\max}}g_{{\cal{N}}}(n)e^{-\frac{E_{{\rm{FM}}}-\frac{N}{2}h-n(h_1-h)}{T}}
\nonumber\\
=
e^{-\frac{E_{{\rm{FM}}}-\frac{N}{2}h}{T}}
\sum_{n=0}^{n_{\max}}g_{{\cal{N}}}(n) z^n,
\;\;\;
z=e^{\frac{h_1-h}{T}},
\;\;\;
h_1=\varepsilon,
\end{eqnarray}
where the quantity $g_{{\cal{N}}}(n)$ 
represents the degeneracy of the ground-state manifold of $n$ magnons in a system with $\cal{N}$ traps.
For the spin models it is easy to obtain
(see Ref.~\onlinecite{localized_magnons_b})
\begin{eqnarray}
\label{3.02}
g_{{\cal{N}}}(n)=2^n{\cal{C}}_{\cal{N}}^n \; , 
\; 
{\cal{C}}_{\cal{N}}^n =\frac{{\cal{N}}!}{n!({\cal{N}}-n)!}
\end{eqnarray}
(double-tetrahedra chain)
and
\begin{eqnarray}
\label{3.03}
g_{{\cal{N}}}(n)=2^n{\cal{Z}}_{{\rm{hd}}}(n,{\cal{N}})
\end{eqnarray}
(frustrated triangular tube),
where ${\cal{Z}}_{{\rm{hd}}}(n,{\cal{N}})$ 
is the canonical partition function of $n$ hard dimers on a periodic chain of ${\cal{N}}$ sites.
The factor $2^n$ in the above expressions stems from the extra degeneracy due to the chirality degrees of freedom.
After substitution of $g_{\cal{N}}(n)$ from Eq.~(\ref{3.02}) or Eq.~(\ref{3.03}) into Eq.~(\ref{3.01}) 
one obtains the free energy
\begin{eqnarray}
\label{3.04}
\frac{F_{{\rm{lm}}}(T,h,N)}{{\cal{N}}}
=
\frac{E_{{\rm{FM}}}}{{\cal{N}}}-\frac{N}{2{\cal{N}}}h-T\ln\left(1+2z\right)
\end{eqnarray}
(double-tetrahedra chain)
or
\begin{eqnarray}
\label{3.05}
\frac{F_{{\rm{lm}}}(T,h,N)}{{\cal{N}}}
=
\frac{E_{{\rm{FM}}}}{{\cal{N}}}-\frac{N}{2{\cal{N}}}h-T\frac{\ln\left(\lambda_1^{{\cal{N}}}+\lambda_2^{{\cal{N}}}\right)}{{\cal{N}}},
\nonumber\\
\lambda_{1,2}=\frac{1}{2}\pm\sqrt{\frac{1}{4}+2z} \qquad \qquad \qquad 
\end{eqnarray}
(frustrated triangular tube). 
At low temperatures and for magnetic fields $h$ around the saturation field $h_1=\varepsilon$  
the contribution of localized states is dominating. 
Hence, 
$F_{{\rm{lm}}}(T,h,N)$ given in Eqs.~(\ref{3.04}) and (\ref{3.05})
yields a good description of the low-temperature physics near the saturation field of the full spin model.

Analogously, 
the contribution of the localized eigenstates to the grand-canonical partition function of the electron models 
is given by 
\begin{eqnarray}
\label{3.06}
\Xi_{{\rm{le}}}(T,\mu,N)=\sum_{n=0}^{n_{\max}}g_{{\cal{N}}}(n)e^{-\frac{(-\varepsilon+\mu)n}{T}}
\nonumber\\
=\sum_{n=0}^{n_{\max}}g_{{\cal{N}}}(n) z^n,
\;\;\;
z=e^{\frac{\varepsilon-\mu}{T}}.
\end{eqnarray}
To avoid the calculation of $g_{{\cal{N}}}(n)$ 
one may rewrite Eq.~(\ref{3.06}) as a sum over occupation numbers of each cell 
taking into account 
(i) that the cells are independent 
and 
(ii) that each cell may contain 0, 1, or 2 electrons 
having the degeneracy of the ground states $g_{1}(0)=1$, $g_{1}(1)=4$, or $g_{1}(2)=3$, respectively,
see Ref.~\onlinecite{localized_electrons}.
Thus, we have
\begin{eqnarray}
\label{3.07}
&& \Xi_{{\rm{le}}}(T,\mu,N)=\nonumber \\ 
&& \quad \sum_{n_1=0,1,2}\ldots\sum_{n_{\cal{N}}=0,1,2}
g_{1}(n_1)\ldots g_{1}(n_{{\cal{N}}})
z^{n_1+\ldots+n_{{\cal{N}}}}\nonumber \\
&&\qquad  =\left[\sum_{n=0,1,2}g_{1}(n)z^{n}\right]^{\cal{N}}
=\left(1+4z+3z^{2}\right)^{\cal{N}}
\end{eqnarray}
for both lattices, the double-tetrahedra chain and the frustrated triangular tube.
Eq.~(\ref{3.07}) immediately yields the required grand-thermodynamical potential
\begin{eqnarray}
\label{3.08}
\frac{\Omega_{{\rm{le}}}(T,\mu,N)}{{\cal{N}}}
=-T\ln\left(1+4z+3z^2\right).
\end{eqnarray} 
Again, 
at low temperatures and for chemical potentials $\mu$ around $\mu_0=\varepsilon$  
the contribution of localized states is dominating, 
and  $\Omega_{{\rm{le}}}(T,\mu,N)$ given in Eq.~(\ref{3.08})
yields a good description of the low-temperature physics of the full electron model.

We mention that the obtained formulas for the free energy and the grand-thermodynamical potential 
are similar (but not identical) to those derived in previous papers,\cite{localized_magnons_b,localized_electrons}
the deviations from the previously derived equations are related to the chirality degrees of freedom. 

Let us briefly discuss the low-temperature thermodynamics 
as it follows from Eqs.~(\ref{3.04}), (\ref{3.05}), and (\ref{3.08}).
The main low-temperature features of the spin models for $h$ around $h_1$ are as follows:
(i) a jump in the ground-state magnetization curve at the saturation field $h_1$
with a preceding wide plateau, 
(ii) a nonzero residual ground-state entropy at the saturation field $h_1$, 
(iii) a low-temperature peak in the specific heat, which moves to $T=0$ as $h$ approaches $h_1$.
Correspondingly, 
for the electron models we have: 
(i) a zero-temperature jump in the averaged number of electrons as a function of the chemical potential at $\mu=\mu_0$, 
(ii) a nonzero residual ground-state entropy for $\mu=\mu_0$ 
(or as a function of the electron concentration $c=n/{\cal{N}}$ for $c\le 2$), 
(iii) a low-temperature peak in the grand-canonical specific heat $C(T,\mu,N)$, 
but a vanishing low-temperature canonical specific heat $C(T,n,N)=0$ for $n\le n_{\max}$ 
(see also the discussion in Ref.~\onlinecite{dj}).

\begin{figure}
\begin{center}
\includegraphics[clip=on,width=77mm,angle=0]{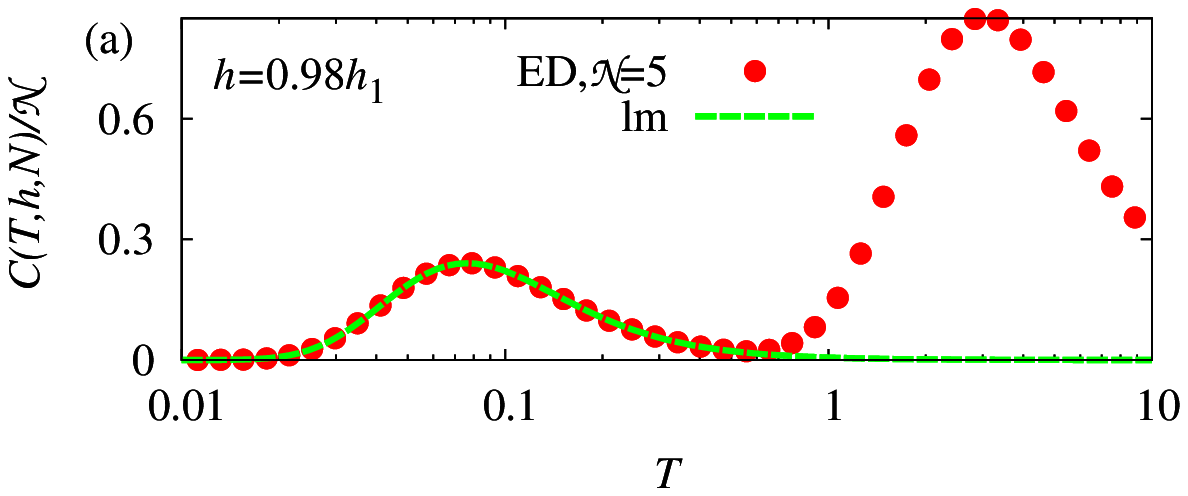}\\
\includegraphics[clip=on,width=77mm,angle=0]{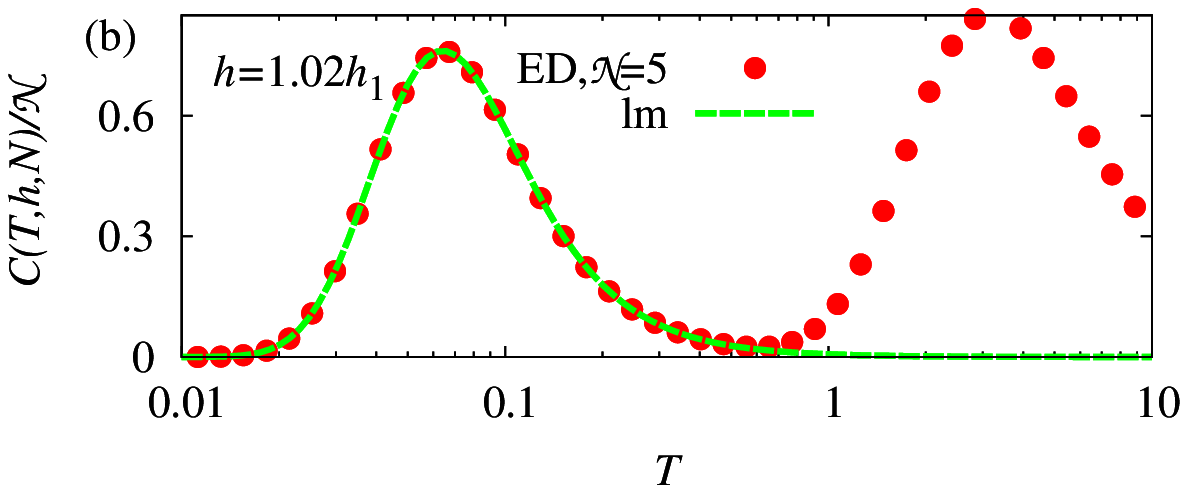}\\
\includegraphics[clip=on,width=77mm,angle=0]{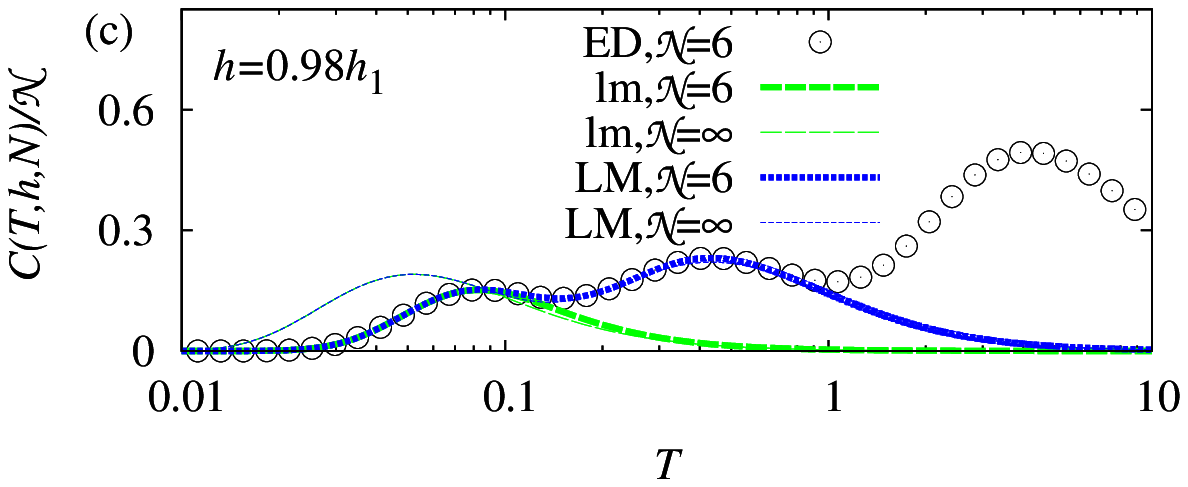}\\
\includegraphics[clip=on,width=77mm,angle=0]{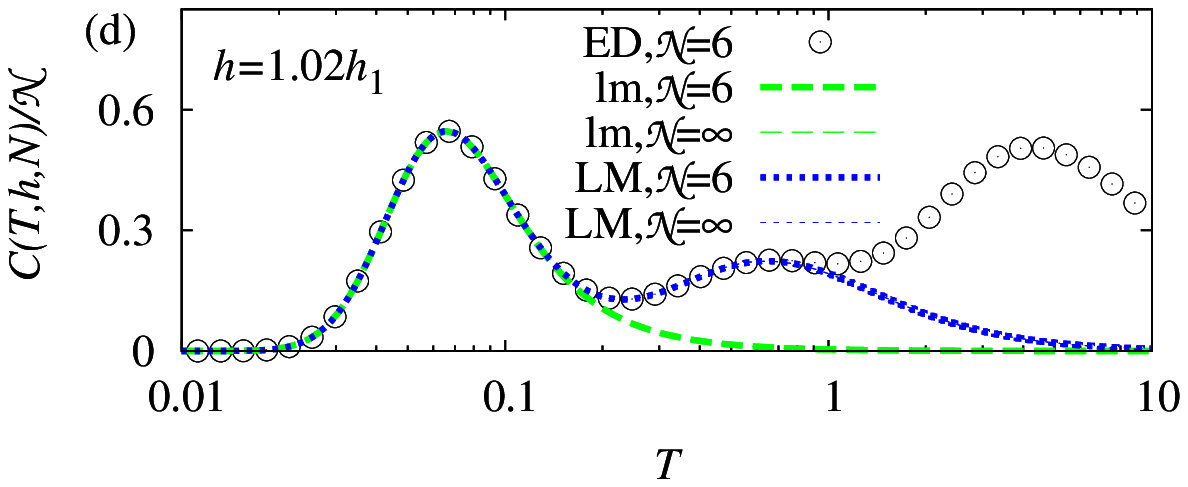}
\caption
{(Color online)
Specific heat $C(T,h,N)/{\cal{N}}$ vs temperature $T$ for $h$ around $h_1$ 
for 
[(a) and (b)] the $N=20$ double-tetrahedra spin chain
and 
[(c) and (d)] the $N=18$ frustrated triangular spin tube.
$J_1=1$, $J_2=5$;
symbols correspond to exact-diagonalization data,
dashed lines correspond to independent localized-magnon predictions 
derived from Eqs.~(\ref{3.04}) and (\ref{3.05})  
(abbreviation lm),
dotted lines correspond to interacting localized-magnon predictions 
derived form Eq.~(\ref{4.02}) 
(abbreviation LM, for the frustrated triangular tube only).
The thin dashed and thin dotted lines in panels (c) and (d)
correspond to localized-magnon predictions in the limit ${\cal{N}}\to\infty$.
Note that some curves in panels (c) and (d) practically coincide.
Namely in panel (c),
thin dashed and thin dotted lines are indistinguishable at low temperatures,
whereas at higher temperatures dashed and thin dashed (dotted and thin dotted) lines coincide.
In panel (d) dashed and thin dashed (dotted and thin dotted) lines cannot be distinguished.}
\label{fig02}
\end{center}
\end{figure}

The temperature dependence of the specific heat of the spin models is shown in Fig.~\ref{fig02}. 
By comparison of the hard-core data obtained from Eqs.~(\ref{3.04}), (\ref{3.05}) 
with exact-diagonalization data of the full spin model 
we can estimate the range of validity of the hard-core description.
The low-temperature parts of the curves for the specific heat 
obtained from exact diagonalization (symbols) and from the hard-core description (dashed lines)
coincide at least up to $T=0.15$. 
For both Heisenberg chains the specific heat shows 
in addition to the typical high-temperature maximum around $T \sim J_2$
a low-temperature maximum 
which is well described by the hard-core models. 
This low-temperature maximum can be ascribed to an extra low-energy scale set by the localized eigenstates.
Interestingly, 
for the frustrated triangular tube there is even a third maximum 
which can be related to a third energy scale set by another class of highly degenerate eigenstates, 
the so-called interacting localized-magnon states,
which will be discussed in the next section.

There are no finite-size effects in the hard-monomer description (\ref{3.04}).
To illustrate the finite-size dependence inherent in the hard-dimer description (\ref{3.05}), 
we compare in panels (c) and (d) of Fig.~\ref{fig02} results for finite $\cal N$ (dashed lines)
with data for ${\cal{N}}\to\infty$ (thin dashed lines). 
Clearly, 
finite-size effects are obvious only at low temperatures for $h$ below the saturation field $h_1$.

The high degeneracy of the localized eigenstates leads to a residual ground-state entropy 
given by 
$S(T=0,h=h_1,N)/{\cal{N}}=\ln 3\approx 1.099$ (double-tetrahedra spin chain)
and 
$S(T=0,h=h_1,N)/{\cal{N}}=\ln 2\approx 0.693$ (frustrated triangular spin tube).
Due to the chirality these numbers exceed the corresponding results
reported in Ref.~\onlinecite{localized_magnons_b}b 
for the standard diamond spin chain $S(T=0,h=h_1,N)/{\cal{N}}=\ln 2\approx 0.693$
and
the frustrated two-leg spin ladder $S(T=0,h=h_1,N)/{\cal{N}}=\ln[(1+\sqrt{5})/2] \approx 0.481$.

\begin{figure}
\begin{center}
\includegraphics[clip=on,width=77mm,angle=0]{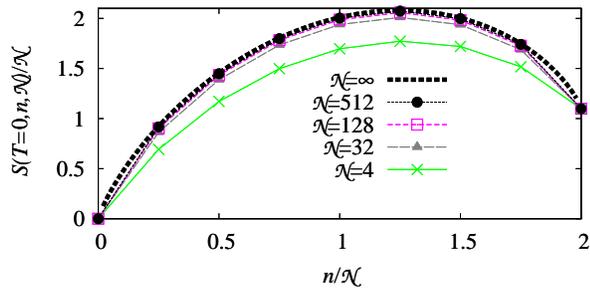}
\caption
{(Color online)
Residual ground-state entropy $S(T=0,n,{\cal{N}})/{\cal{N}}$ 
vs 
electron concentration $c=n/{\cal{N}}$ 
for the Hubbard model with $t_2>2t_1$
on the double-tetrahedra chain and the frustrated triangular tube 
(the data for both systems are identical).
Note that the curves becomes indistinguishable in the scale of the figure as ${\cal{N}}$ increases.}
\label{fig03}
\end{center}
\end{figure}

For the Hubbard model we have calculated curves for the grand-canonical specific heat 
similar to those for the Heisenberg model, 
which for the sake of brevity are not shown here. 
They also exhibit an additional low-temperature maximum for $\mu \lesssim \mu_0$ and $\mu \gtrsim \mu_0$ 
which is well described by the localized eigenstates, 
see also Ref.~\onlinecite{localized_electrons}.
The residual ground-state entropy
$\lim_{T\to 0}S(T,n,N)=\ln g_{{\cal{N}}}(n)$
as a function of the electron concentration $c=n/{\cal{N}}$ 
for the Hubbard model
is shown  in Fig.~\ref{fig03}.
According to Eq.~(\ref{3.06}) 
for finite systems one has 
$n!g_{\cal{N}}(n)={d^n \Xi_{{\rm{le}}}(z,N)/dz^n}\vert_{z=0}$. 
Using Eq.~(\ref{3.07}) one obtains $g_{{\cal{N}}}(n)$ for (small) finite ${\cal{N}}$. 
For ${\cal{N}}=4$ the analytical predictions which follow from Eqs.~(\ref{3.07}), (\ref{3.08})
coincide with exact-diagonalization data for the full Hubbard model,
e.g.,
$g_{4}(n)=1,16,108,400,886,1200,972,432,81$
for
$n=0,1,2,3,4,5,6,7,8$.
Using  Eq.~(\ref{3.08}) by means of standard relations of statistical mechanics 
we find for the residual ground-state entropy in the thermodynamic limit 
$S(T=0,n,N)/{\cal{N}}=\ln(1+4z+3z^2)-(4z\ln z+6z^2\ln z)/(1+4z+3z^2)$
with
$z=[2(1-c)-\sqrt{c^2-2c+4}]/[3(c-2)]$, $c=n/{\cal{N}}$, see the dotted curve in Fig.~\ref{fig03}.
This quantity reaches its maximum, $3\ln 2\approx 2.079$, at $c=n/{\cal{N}}=5/4$.
Moreover, it equals $\ln 3\approx 1.099$ at $c=n/{\cal{N}}=2$.
Obviously,
the chirality leads to an increase of the residual ground-state entropy.
While for the Hubbard model on the diamond chain and the frustrated two-leg ladder
$S(T=0,\mu=\mu_0,N)/{\cal{N}}=\ln 3\approx 1.099$ (see Ref.~\onlinecite{localized_electrons}b),
for the considered lattices
$S(T=0,\mu=\mu_0,N)/{\cal{N}}=3\ln 2\approx 2.079$.

\section{Beyond independent localized magnons}
\label{sec4}
\setcounter{equation}{0}

For the frustrated-tube spin model we can extend the hard-dimer description 
taking into account additional low-energy states 
following the lines described in Ref.~\onlinecite{labi}. 
If one allows the occupation of neighboring traps by localized magnons
(i.e., relaxing the hard-dimer rule) 
one has also an eigenstate of the spin Hamiltonian, 
however with a higher energy.
More precisely,
if two localized magnons become neighbors 
the energy increases by $J_1$.
Importantly,
these localized-magnon states are also highly degenerate 
and they are the lowest excitations above the independent localized-magnon ground states 
for $S^z=N/2,\ldots,N/2-{\cal{N}}/2$
if $J_2>J_2^c$
(strong-coupling regime).
Based on finite-size calculations for $N=18,\dots,72$ 
we estimate $J_2^c/J_1\approx 2.68>2$. 
These additional eigenstates can be described as interacting localized-magnon states, 
where the repulsive interaction $V=J_1$ is responsible for the energy increase 
with respect to the independent localized-magnon states.
Taking  into account this finite repulsion $V$ 
in the partition function of the lattice-gas model with nearest-neighbor interaction
we have then [instead of Eq.~(\ref{3.01})]
\begin{eqnarray}
\label{4.01}
Z_{{\rm{LM}}}(T,h,N)
=
e^{-\frac{E_{{\rm{FM}}}-\frac{N}{2}h}{T}}
\sum_{n_1=0,1}\ldots\sum_{n_{{\cal{N}}}=0,1}
g_{1}(n_1)
\nonumber\\
\times
\ldots g_{1}(n_{\cal{N}})
z^{n_1+\ldots+n_{\cal{N}}}
e^{-\frac{V(n_1n_2+n_2n_3+\ldots+n_{\cal{N}}n_1)}{T}} \quad
\end{eqnarray}
with $g_{1}(0)=1$, $g_{1}(1)=2$ and $z$ is defined in Eq.~(\ref{3.01}).
In Eq.~(\ref{4.01})
we have used a representation in terms of the cell occupation numbers $n_i$,
cf. Eq.~(\ref{3.07}).
Obviously,  Eq.~(\ref{3.01}) is obtained from   Eq.~(\ref{4.01}) for $V\to \infty$.
Evaluating the sums in Eq.~(\ref{4.01}) by means of the transfer-matrix method 
we arrive at the following result for the free energy
\begin{eqnarray}
\label{4.02}
\frac{F_{{\rm{LM}}}(T,h,N)}{{\cal{N}}}
=\frac{E_{{\rm{FM}}}}{\cal{N}}-\frac{N}{2{\cal{N}}}h
-T\frac{\ln\left(\lambda_1^{\cal{N}}+\lambda_2^{\cal{N}}\right)}{\cal{N}},
\nonumber\\
\lambda_{1,2}=\frac{1}{2}+ze^{-\frac{V}{T}}
\pm\sqrt{\frac{1}{4}+2z-ze^{-\frac{V}{T}}+z^2e^{-\frac{2V}{T}}}.
\end{eqnarray}
Note that the lattice-gas model with finite repulsion (\ref{4.01}), (\ref{4.02}) 
takes into account $3^{{\cal{N}}}\approx 1.442^N$ states\cite{footnote1} 
of the $2^N$ eigenstates of the initial quantum spin model (\ref{2.01}),
whereas hard-dimer model (\ref{3.01}), (\ref{3.03}), (\ref{3.05}) 
has only $2^{{\cal{N}}}\approx 1.260^N$ states.\cite{footnote2}
On the other hand, 
the hard-monomer model (\ref{3.01}), (\ref{3.02}), (\ref{3.04}) 
has $3^{{\cal{N}}}\approx 1.316^N$ states.\cite{footnote1}

The specific heat derived from Eq.~(\ref{4.02}) is plotted in Fig.~\ref{fig02}, panels (c) and (d), dotted and thin dotted lines.
Indeed, 
the inclusion of the interacting localized-magnon states leads to a significant improvement of the lattice-gas description.  
The lattice-gas model with finite repulsion (\ref{4.01}), (\ref{4.02}) covers the thermodynamics of the full spin model 
at least up to $T=0.9$ for $h \sim h_1$ including the two maxima 
below the typical high-temperature maximum around $T \sim J_2$.
Again finite-size effects are more important at low temperatures for $h$ below the saturation field $h_1$,
see thin dotted lines in panels (c) and (d) of Fig.~\ref{fig02}.

\begin{figure}
\begin{center}
\includegraphics[clip=on,width=77mm,angle=0]{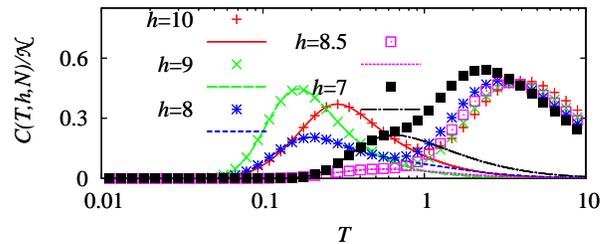}
\caption
{(Color online)
Specific heat $C(T,h,N)/{\cal{N}}$ vs temperature $T$ 
for finite frustrated triangular spin tubes of $N=18$ sites (${\cal{N}}=6$ cells) with $J_1=1$, $J_2=5$ ($h_1=10.5$)
for various magnetic fields down to values below $h_2=8.5$ 
(lines -- lattice-gas model with finite repulsion, 
symbols -- exact-diagonalization data for the full spin model).}
\label{fig04}
\end{center}
\end{figure}

The interacting localized-magnon states being excitations for $S^z=N/2,\ldots,N/2-{\cal{N}}/2$ 
can become ground states for smaller values of $S^z$.
The ground-state magnetization curve $\langle M\rangle=2S^z/N$ vs $h$ for the frustrated triangular spin tube 
presented in Ref.~\onlinecite{andreas} 
exhibits two plateaus 
at $\langle M\rangle=2/3$ for $h_2<h<h_1$  
and 
at $\langle M\rangle=1/3$ for $h_3<h<h_2$ 
and two jumps at $h=h_1$ and $h=h_2$, 
where $h_1=3J_1+3J_2/2$, $h_2=J_1+3J_2/2$, and $h_3=2J_1$.
In the language of localized magnons 
the $2/3$-plateau corresponds to the maximum filling with $n=n_{\max}={\cal N}/2$ independent localized magnons 
(i.e., every second trap is occupied and $S^z=N/2-{\cal N}/2=N/3$).
If $n_{\max} < n \le 2n_{\max}$
(i.e., $N/2-{\cal{N}} \le S^z < N/2-{\cal{N}}/2$),
the ground states in the strong-coupling regime are obtained by filling the remaining empty cells 
(i.e., the hard-dimer rule is relaxed)  
thus having interacting localized-magnon states as ground states.
Then the very broad  $1/3$-plateau 
corresponds to the complete filling of all cells with $n=2n_{\max}={\cal N}$ localized magnons. 
Hence,
with the improved effective theory given in Eq.~(\ref{4.02}) 
we can provide an accurate description 
of the low-temperature physics of the frustrated triangular spin tube in the strong-coupling regime  
not only near the saturation field $h_1$ up to quite large temperatures 
as shown in  Fig.~\ref{fig02}, 
but also for much lower magnetic fields in the entire region of the  $2/3$-plateau  
and even for fields  within the $1/3$-plateau being not too far from $h_2$,
see Fig.~\ref{fig04}.
It might be interesting to recall
that the lattice-gas model with finite repulsion 
provides similar description of the frustrated two-leg spin ladder 
around both characteristic fields $h_1$ and $h_2$.\cite{labi}
This is not the case for the frustrated triangular spin tube
because of the chirality degrees of freedom,
compare the results for $h=10$ and $h=9$ in Fig.~\ref{fig04}.

\section{Lifting the degeneracy due to chirality degrees of freedom}
\label{sec5}
\setcounter{equation}{0}

As discussed above 
the chirality degrees of freedom lead to an extra degeneracy 
of the independent localized-magnon or localized-electron ground states 
in the subspace with $n\le n_{\max}$ magnons or electrons.
Moreover, 
the interacting localized-magnon states discussed for the frustrated triangular spin tube 
also carry this extra degree of freedom.
This degeneracy of the eigenstates owing to the chirality may be lifted by a small symmetry-breaking perturbation.
As a rule, 
perturbations of ideal model Hamiltonians may lead to a more realistic description of real systems.
We consider here separately for the spin systems 
the case of a Zeeman-like perturbation 
(which corresponds to a Dzyaloshinskii-Moriya interaction between neighboring spins in a triangular trap),
see Sec.~\ref{sec5a},
and 
the case of an $XX$-like perturbation acting on (pseudo)spin variables representing chiralities  
(which corresponds to a four-site interaction between spin pairs in neighboring traps), 
see Sec.~\ref{sec5c}. 
For the Hubbard model 
appropriate perturbations correspond to 
a magnetic field perpendicular to the triangular traps
and 
a four-site electron-electron interaction, 
see Sec.~\ref{sec5b} and Sec.~\ref{sec5c}.

\subsection{Dzyaloshinskii-Moriya interaction}
\label{sec5a}

We consider the spin system in the subspaces with $n\le n_{\max}$ magnons. 
The ground state has the energy $E_{\rm{FM}}-n\varepsilon$ ($h=0$)
and the degeneracies $g_{\cal{N}}(n)$ are given in Eqs.~(\ref{3.02}) or (\ref{3.03}).
The factor $2^n$ present in these formulas for $g_{\cal{N}}(n)$ is due to the chirality of localized magnons.
We introduce the (pseudo)spin-1/2 operators 
\begin{eqnarray}
\label{5.01}
\tau_m^z
=\frac{1}{2}\left(\vert +\rangle_m\langle +\vert_m-\vert -\rangle_m\langle -\vert_m\right)
=\frac{1}{2}\chi_m,
\end{eqnarray}
where $\chi_m=(i/\sqrt{3})(s_{m,1}^+s_{m,2}^--s_{m,1}^-s_{m,2}^+
+ s_{m,2}^+s_{m,3}^--s_{m,2}^-s_{m,3}^+ + s_{m,3}^+s_{m,1}^--s_{m,3}^-s_{m,1}^+)$
is the chirality operator, see Eq.~(\ref{2.09}) and the discussion below this equation.
We now add to the spin Hamiltonian (\ref{2.01}) a small perturbation
\begin{eqnarray}
\label{5.02}
H_{\rm{s}}^{(1)}
&=&\frac{i\epsilon^{(1)}}{2\sqrt{3}}\sum_{m}\left(s_{m,1}^+s_{m,2}^--s_{m,1}^-s_{m,2}^+
\right.
\nonumber\\
&&\left.
+s_{m,2}^+s_{m,3}^--s_{m,2}^-s_{m,3}^+
+s_{m,3}^+s_{m,1}^--s_{m,3}^-s_{m,1}^+\right)
\nonumber\\
&=&
D\sum_{m}\left(s_{m,1}^xs_{m,2}^y-s_{m,1}^ys_{m,2}^x
\right.
\nonumber\\
&&\left.
+s_{m,2}^xs_{m,3}^y-s_{m,2}^ys_{m,3}^x
+s_{m,3}^xs_{m,1}^y-s_{m,3}^ys_{m,1}^x\right), 
\nonumber
\\
\end{eqnarray}
where the last expression in Eq.~(\ref{5.02}) corresponds to a Dzyaloshinskii-Moriya interaction
${\vec D}=D{\vec e_z}$, $D=\epsilon^{(1)}/\sqrt{3}$ 
between neighboring spins within each triangular trap.
Note that the perturbation $H_{\rm{s}}^{(1)}$ commutes with $S^z$.
According to the discussion of the chirality operator (\ref{2.09}) in Sec.~\ref{sec2} 
it is obvious that the localized states are also eigenstates of the perturbation Hamiltonian (\ref{5.02}).
The set of $2^n$ degenerate ground states in the unperturbed system
belonging to one particular spatial configuration of $n$ magnons placed in a certain (allowed) set of traps 
now splits into $n+1$ subsets of levels.
The subsets are characterized by the magnon numbers $n_+$ and $n_-$ ($n_++n_-=n$), 
belonging to the  chirality indeces $+$ and $-$, respectively.
There are $n!/(n_+!n_-!)$ degenerate states 
in the subset with energy $E_{\rm{FM}}-n\varepsilon+(n_+-n_-)\epsilon^{(1)}/2$.
The effective Hamiltonian 
acting in the subspace of the $g_{\cal{N}}(n)$ former $n$-magnon ground states of the unperturbed system 
reads
\begin{eqnarray}
\label{5.03}
{\cal{H}}_{\rm{s}}+{\cal{H}}_{\rm{s}}^{(1)}=E_{\rm{FM}}-n\varepsilon + \epsilon^{(1)}\sum_m\tau_m^z,
\end{eqnarray}
where the sum runs over the $n$ occupied traps only.

We consider now the partition function of the spin model with the Hamiltonian $H_{\rm{s}}+H_{\rm{s}}^{(1)}$,
Eqs.~(\ref{2.01}) and (\ref{5.02}),
at low  temperatures and $h$ close to $h_1$.
The dominant contribution to the partition function of the spin system comes from the low-energy degrees of freedom,
which are governed by the Hamiltonian (\ref{5.03}).
Therefore the partition function is given by Eq.~(\ref{3.01}) replacing $g_{\cal{N}}(n)$ by 
${\tilde g}_{\cal{N}}(n)=\{2\cosh[\epsilon^{(1)}/(2T)]\}^n{\cal{C}}_{\cal{N}}^n$
(double-tetrahedra chain)
and
${\tilde g}_{\cal{N}}(n)=\{2\cosh[\epsilon^{(1)}/(2T)]\}^n{\cal{Z}}_{\rm{hd}}(n,{\cal{N}})$
(frustrated triangular tube),
cf. Eqs.~(\ref{3.02}) and (\ref{3.03}).
As a result, 
the free energy reads
\begin{eqnarray}
\label{5.04}
\frac{F_{{\rm{lm}}}(T,h,N)}{{\cal{N}}}
=
\frac{E_{{\rm{FM}}}}{{\cal{N}}}-\frac{N}{2{\cal{N}}}h
-T\ln\left(1+2\cosh\frac{\epsilon^{(1)}}{2T}z\right)  
\end{eqnarray}
(double-tetrahedra chain)
and
\begin{eqnarray}
\label{5.05}
\frac{F_{{\rm{lm}}}(T,h,N)}{{\cal{N}}}
&=&
\frac{E_{{\rm{FM}}}}{{\cal{N}}}-\frac{N}{2{\cal{N}}}h-T\frac{\ln\left(\lambda_1^{{\cal{N}}}+\lambda_2^{{\cal{N}}}\right)}{{\cal{N}}},
\nonumber\\
\lambda_{1,2}&=&\frac{1}{2}\pm\sqrt{\frac{1}{4}+2\cosh\frac{\epsilon^{(1)}}{2T}z}   
\end{eqnarray}
(frustrated triangular tube),
cf. Eqs.~(\ref{3.04}) and (\ref{3.05}).
Similar to Sec. \ref{sec4},
we can take into account interacting localized-magnon states also for the perturbed frustrated triangular tube 
described by the Hamiltonian $H_{\rm{s}}+H_{\rm{s}}^{(1)}$. 
The improved free energy $F_{{\rm{LM}}}(T,h,N)$ is then given by  Eq.~(\ref{4.02}), 
where $2z$ has to be substituted by $2\cosh[\epsilon^{(1)}/(2T)] z$.

\begin{figure}
\begin{center}
\includegraphics[clip=on,width=77mm,angle=0]{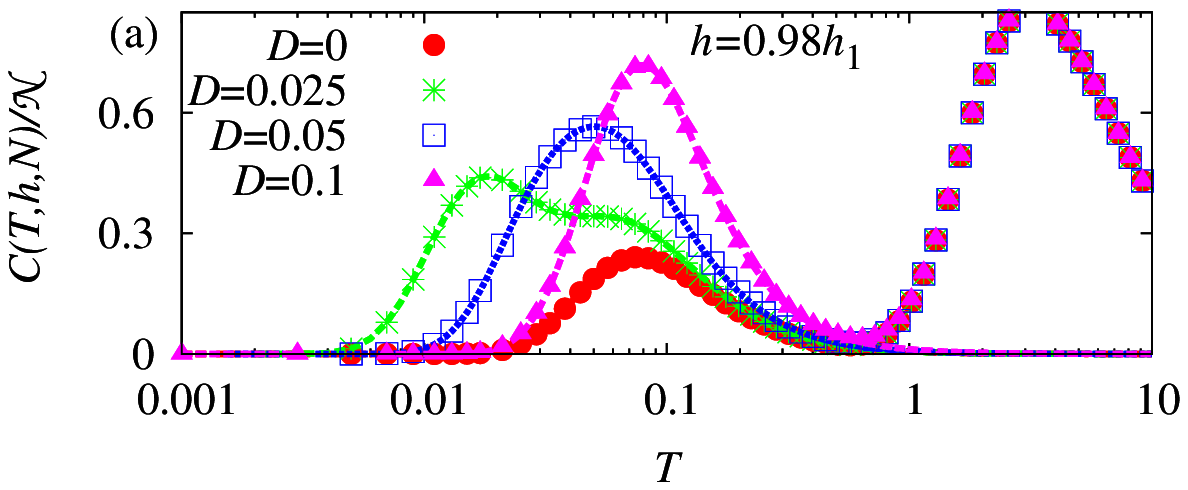}\\
\includegraphics[clip=on,width=77mm,angle=0]{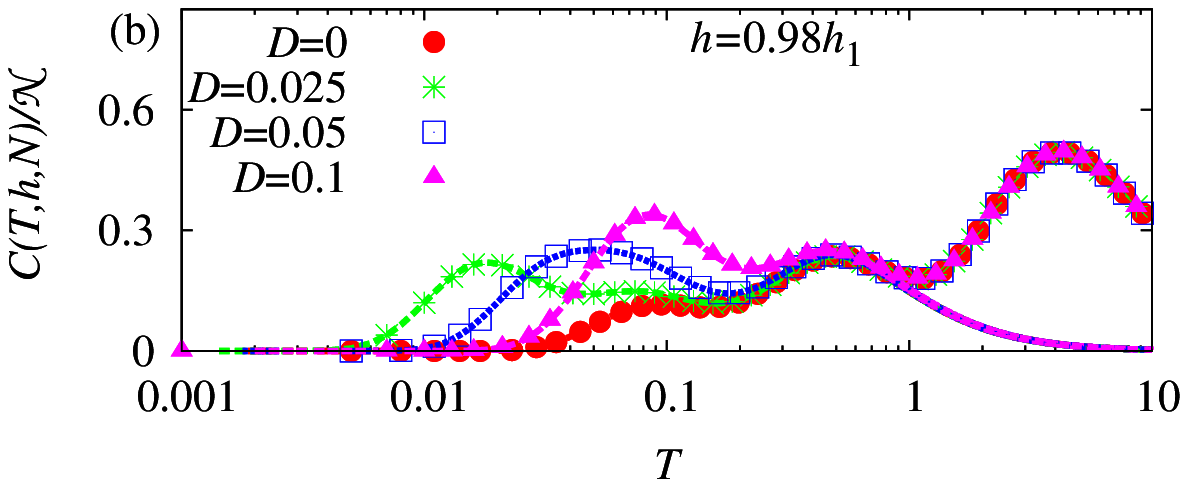}
\caption
{(Color online)
Specific heat $C(T,h,N)/{\cal{N}}$ vs temperature $T$ for $h=0.98h_1$ for the spin model (\ref{2.01}), (\ref{5.02}) on 
(a) the double-tetrahedra chain 
and 
(b) the frustrated triangular tube.
Exact-diagonalization data (symbols)
versus
analytical predictions (lines)
according to Eq.~(\ref{5.04}) (double-tetrahedra chain) 
and improved Eq.~(\ref{5.05}) (frustrated triangular tube)
for finite systems of ${\cal{N}}=4$ cells 
with $J_1=1$, $J_2=5$, and $D=0,\,0.025,\,0.05,\,0.1$.}
\label{fig05}
\end{center}
\end{figure}

In Fig.~\ref{fig05} we compare exact-diagonalization data for perturbed spin systems
with predictions based on Eq.~(\ref{5.04}) and improved Eq.~(\ref{5.05}). 
Small perturbations lead to splitting of the ground-state levels 
and therefore to arising of one more low-energy scale.
As a result, 
low-temperature features close to $h_1$ are more subtle.
Thus temperature profiles of the specific heat  for the spin systems show more tiny features 
which can be seen in Fig.~\ref{fig05}.
For a special set of parameters 
the temperature dependence $C$ vs $T$ may exhibit even three (four) maxima 
for the double-tetrahedra chain (frustrated triangular tube).
The low-temperature maxima are excellently described within the effective low-energy theory,
see Fig.~\ref{fig05}.

\subsection{Electrons in a magnetic field}
\label{sec5b}

In analogy to the above discussion for perturbed spin models, 
we consider the case of $n\le n_{\max}$ Hubbard electrons (\ref{2.02})
with a perturbation
\begin{eqnarray}
\label{5.06}
H_{\rm{e}}^{(1)}
=
\sum_{\sigma=\uparrow,\downarrow}H_{0,\sigma}^{(1)},
\nonumber\\
H_{0,\sigma}^{(1)}
=
ig\sum_{m}
\left(c_{m,1,\sigma}^{\dagger}c_{m,2,\sigma}+c_{m,1,\sigma}c_{m,2,\sigma}^{\dagger}
\right.
\nonumber\\
\left.
+c_{m,2,\sigma}^{\dagger}c_{m,3,\sigma}+c_{m,2,\sigma}c_{m,3,\sigma}^{\dagger}
\right.
\nonumber\\
\left.
+c_{m,3,\sigma}^{\dagger}c_{m,1,\sigma}+c_{m,3,\sigma}c_{m,1,\sigma}^{\dagger}
\right),
\end{eqnarray}
where $ig$ is a pure imaginary component of the hopping integral between neighboring sites along the triangular traps.
For the perturbed Hamiltonian the number of electrons remains a conserved quantity 
and the localized states are its eigenstates, 
cf. the chirality operator (\ref{2.10}) in Sec.~\ref{sec2}.
The perturbation considered here corresponds to a magnetic field perpendicular to the triangular trap.
Then the hoping integral $t_{{\bf{i}}{\bf{j}}}$ acquires the Peierls phase factor 
$e^{(2i\pi/\Phi_0)\int_{{\bf{i}}}^{{\bf{j}}}d{\bf{r}}\cdot{\bf{A}}}$, 
where
$\Phi_0=hc/e$ is the flux quantum, 
${\bf{A}}$ is the vector potential of the external magnetic field,
see Refs.~\onlinecite{essler,gula}. 
The effective Hamiltonian 
acting in the subspace of the $g_{\cal{N}}(n)$ former $n$-electron ground states of the unperturbed system 
reads
\begin{eqnarray}
\label{5.07}
{\cal{H}}_{\rm{e}}+{\cal{H}}_{\rm{e}}^{(1)}
=-n\varepsilon + \epsilon^{(1)}\sum_m\tau_m^z,
\end{eqnarray}
where $\epsilon^{(1)}=-2\sqrt{3}g$
and the sum runs over the $n$ occupied traps only.

We consider now the grand-canonical partition function of the electron models 
with the Hamiltonian $H_{\rm{e}}+H_{\rm{e}}^{(1)}$,
Eqs.~(\ref{2.02}) and (\ref{5.06}),
at low  temperatures and $\mu$ close to $\mu_0=\varepsilon$.
The dominant contribution to the grand-canonical partition function comes from the low-energy states, 
which are governed by the Hamiltonian (\ref{5.07}).
Repeating the arguments which lead to Eq.~(\ref{3.08})
we arrive now at
\begin{eqnarray}
\label{5.08}
\frac{\Omega_{{\rm{le}}}(T,\mu,N)}{{\cal{N}}}
=-T\ln\left(1+4\cosh\frac{\epsilon^{(1)}}{2T}z+3z^2\right),
\end{eqnarray} 
where $z=e^{(\varepsilon-\mu)/T}$, cf. Eqs.~(\ref{3.06}) -- (\ref{3.08}).

We focus on the low-temperature behavior of the entropy of the perturbed electron model. 
We can easily find, using a simple counting of states, the residual ground-state entropy,
namely,
$S(T=0,n,{\cal{N}})=\ln\left(2^n{\cal{C}}_{\cal{N}}^{n}\right)$ 
for $n=0,\ldots,{\cal{N}}$
and
$S(T=0,n,{\cal{N}})=\ln\left(3^{n-{\cal{N}}}2^{2{\cal{N}}-n}{\cal{C}}_{\cal{N}}^{n-{\cal{N}}}\right)$
for $n={\cal{N}},\ldots,2{\cal{N}}$.
In the thermodynamic limit ${\cal{N}}\to\infty$ this gives:
$S(T=0,n,{\cal{N}})/{\cal{N}}=c\ln 2 -c\ln c-(1-c)\ln(1-c)$ for $0\le c\le 1$
and
$S(T=0,n,{\cal{N}})/{\cal{N}}=(c-1)\ln 3+(2-c)\ln 2 -(c-1)\ln (c-1)-(2-c)\ln(2-c)$ for $1\le c\le 2$,
where $c=n/\cal{N}$.
For the special electron concentrations $c=1$ and $c=2$ 
one has  $S(T=0)/{\cal{N}}=\ln 2\approx 0.693$ and $S(T=0)/{\cal{N}}=\ln 3\approx 1.099$, respectively.
For finite temperatures we find from Eq.~(\ref{5.08}) 
$S(T,n,{\cal{N}})/{\cal{N}}
=
\ln\{1+4\cosh[\epsilon^{(1)}/(2T)]z+3z^2\}
-
\{2(\epsilon^{(1)}/T)\sinh[\epsilon^{(1)}/(2T)]z
+4\cosh[\epsilon^{(1)}/(2T)]z\ln z+6z^2\ln z\}
/\{1+4\cosh[\epsilon^{(1)}/(2T)]z+3z^2\}$
with
$z=\{2(1-c)\cosh[\epsilon^{(1)}/(2T)]-\sqrt{4(c-1)^2\cosh^2[\epsilon^{(1)}/(2T)]-3c(c-2)}\}
/[3(c-2)]$.

\begin{figure}
\begin{center}
\includegraphics[clip=on,width=77mm,angle=0]{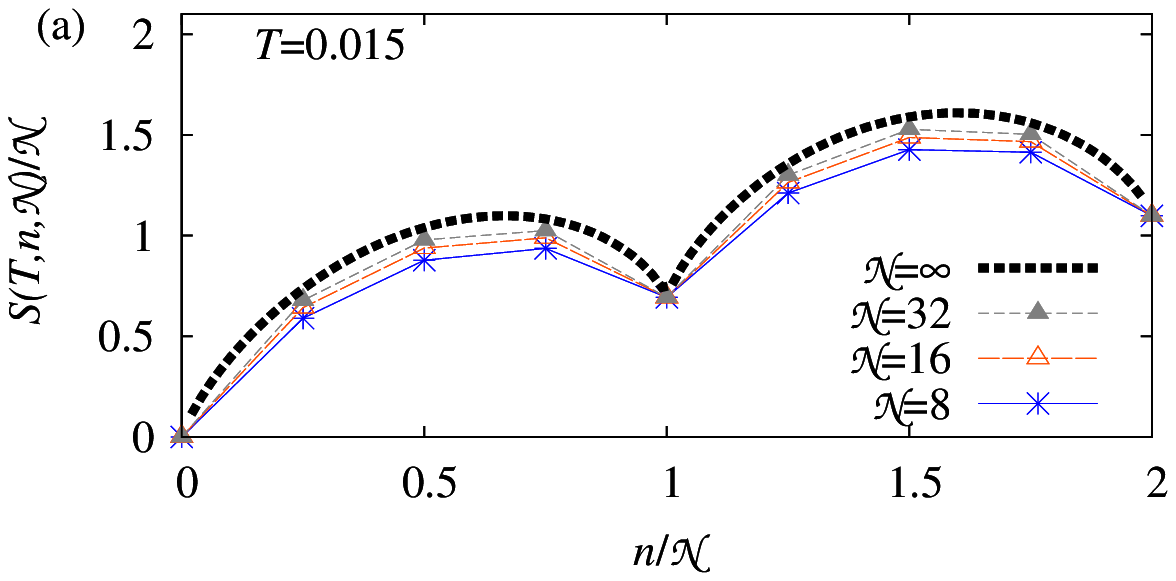}\\
\includegraphics[clip=on,width=77mm,angle=0]{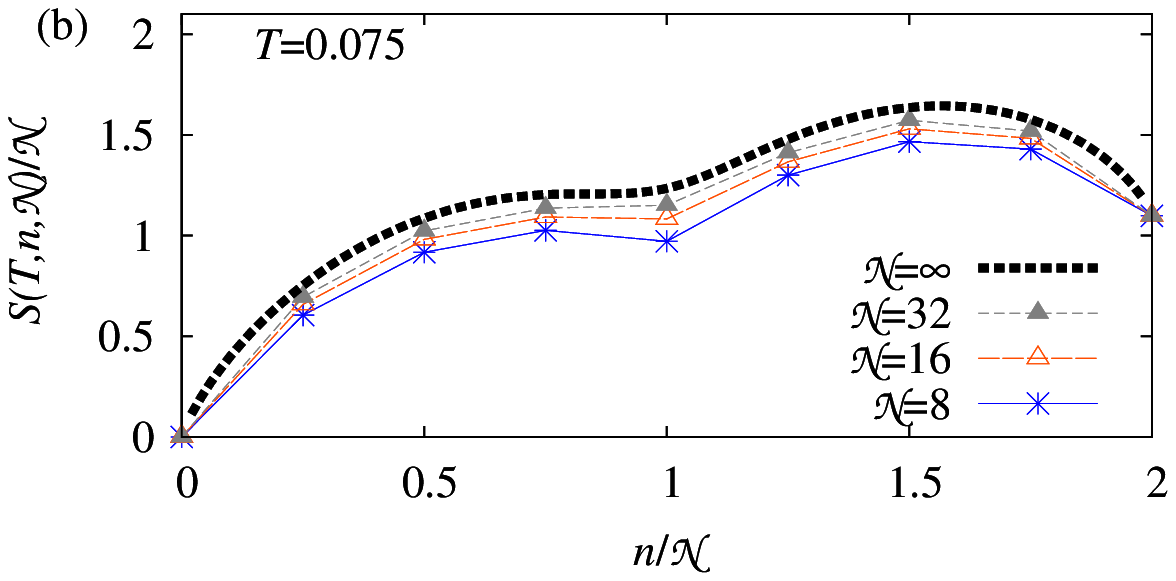}\\
\includegraphics[clip=on,width=77mm,angle=0]{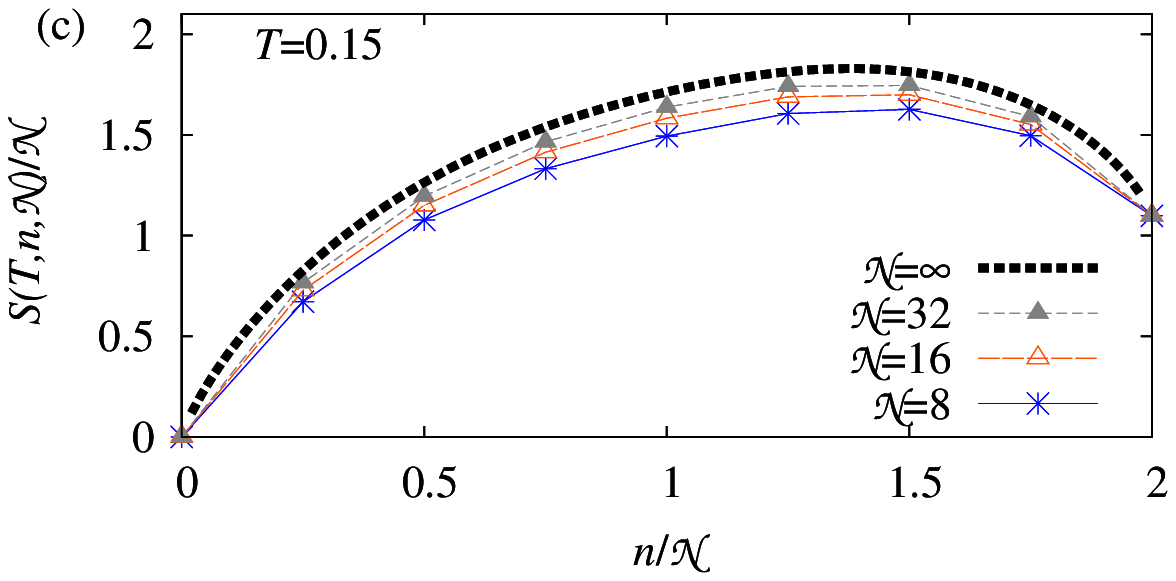}
\caption
{(Color online)
Entropy $S(T,n,{\cal{N}})/{\cal{N}}$ vs electron concentration $c=n/{\cal{N}}$ at low temperatures 
(a) $T=0.015$,
(b) $T=0.075$,
and
(c) $T=0.15$
for the Hubbard model (\ref{2.02}) with $t_2>2t_1$ 
completed by small perturbation (\ref{5.06}) with $g=0.1$
on the double-tetrahedra chain and the frustrated triangular tube.
Analytical predictions are obtained according to Eq.~(\ref{5.08}) 
for systems of ${\cal{N}}=8,\;16,\;32$, and ${\cal{N}}\to\infty$ cells.}
\label{fig06}
\end{center}
\end{figure}

In Fig.~\ref{fig06} we show the entropy $S(T,n,N)$ of the perturbed electron systems obtained by the above given formulas 
versus the electron concentration $c=n/\cal{N}$.
To estimate the region of validity of these results
we compared first  exact-diagonalization data for the grand-canonical specific heat $C(T,\mu,N)$
for finite perturbed electron systems 
(e.g., for ${\cal{N}}=2$, $t_1=1$, $t_2=3$ and $t_2=5$, $g=0.1$, $\mu=0.98\mu_0,\,\mu_0,\,1.02\mu_0$) 
with analytical results for $C(T,\mu,N)$ based on Eq.~(\ref{5.08}). 
(For the sake of brevity we do not show these results explicitely.)  
We find an excellent agreement between these results at least up to $T=0.2$ for $t_2=3$
(and even for higher temperatures for larger values of $t_2$).
For nonzero but low temperatures 
(e.g., $T=0.15$)
$S(T,n,{\cal{N}})/{\cal{N}}$ behaves as in Fig.~\ref{fig03},
see panel (c) of Fig.~\ref{fig06}.
However, at lower temperatures 
(e.g., $T=0.075$ and $T=0.015$)
the smallest energy scale comes into play,
and the entropy changes remaining nonzero in the ground state,
see panels (b) and (a) of Fig.~\ref{fig06}.
In spite of the partial degeneracy lifting due to the perturbation (\ref{5.06}),
the ground states remain hugely degenerate 
and exhibit a nonzero residual ground-state entropy for electron concentrations $0<c=n/{\cal{N}}\le 2$,
see panel (a) of Fig.~\ref{fig06} ($T=0.015$).

\subsection{Interacting pseudospins}
\label{sec5c}

Now we consider the double-tetrahedra spin chain at the $\langle M\rangle=1/2$ plateau 
(i.e., there are $n=n_{\max}=\cal{N}$ magnons in the spin system,
that is the so-called localized-magnon crystal state).
We add a perturbation 
which can be understood as an $XX$ interaction of chirality pseudospins on neighboring trapping cells.
The (pseudo)spin raising and lowering operators are defined as 
\begin{eqnarray}
\label{5.09}
\tau_m^{+}=\vert +\rangle_m\langle -\vert_m,
\;\;\;
\tau_m^{-}=\vert -\rangle_m\langle +\vert_m.
\end{eqnarray}
They can be expressed by bilinear forms in the spin operators $s^-$ and $s^+$ attached to the $m$th cell,
see Eq.~(\ref{2.07}).
Clearly,
$\tau_m^{+}\vert +\rangle_m=0$,
$\tau_m^{+}\vert -\rangle_m=\vert +\rangle_m$,
$\tau_m^{-}\vert +\rangle_m=\vert -\rangle_m$,
$\tau_m^{-}\vert -\rangle_m=0$.
The perturbation
added to the Hamiltonian (\ref{2.01}) reads
\begin{eqnarray}
\label{5.10}
H_{\rm{s}}^{(2)}
=
\frac{\epsilon^{(2)}}{9}\sum_{m}
\left[
\left(s_{m,1}^-+\omega s_{m,2}^-+\omega^2 s_{m,3}^-\right)
\right.
\nonumber\\
\left.
\times
\left(s_{m,1}^++\omega s_{m,2}^++\omega^2 s_{m,3}^+\right)
\right.
\nonumber\\
\left.
\times
\left(s_{m+1,1}^-+\omega^2 s_{m+1,2}^-+\omega s_{m+1,3}^-\right)
\right.
\nonumber\\
\left.
\times
\left(s_{m+1,1}^++\omega^2 s_{m+1,2}^++\omega s_{m+1,3}^+\right)
+{\rm{H.c.}}
\right],
\end{eqnarray}
where the sum runs over all trapping cells and $\omega=e^{2\pi i/3}$.
Obviously, 
that corresponds to certain four-site interactions with the interaction constant
$\epsilon^{(2)}/9$.
The perturbation Hamiltonian $H_{\rm{s}}^{(2)}$ commutes with $S^z$.
Moreover,
after acting on the localized-magnon crystal state 
the perturbation Hamiltonian $H_{\rm{s}}^{(2)}$ changes the chirality indeces only.
The effective Hamiltonian acting in the subspace of the localized-magnon crystal states of the unperturbed system now reads
\begin{eqnarray}
\label{5.11}
{\cal{H}}_{\rm{s}}+{\cal{H}}_{\rm{s}}^{(2)}
=E_{\rm{FM}}-n\varepsilon 
+\epsilon^{(2)}\sum_{m=1}^{{\cal{N}}}\left(\tau_m^{+}\tau_{m+1}^{-}+{\rm{H.c.}}\right).
\end{eqnarray}
Due to the perturbation ${\cal{H}}_{\rm{s}}^{(2)}$
the $2^{{\cal{N}}}$-fold degenerate ground state of the double-tetrahedra spin chain with $n=n_{\max}={\cal{N}}$ magnons
splits into ${\cal{N}}+1$ groups of sublevels.
Moreover, the effective (pseudo)spin-1/2 $XX$ chain (\ref{5.11}) is the exactly solvable model\cite{lieb}
and therefore we immediately obtain for the partition function $Z(T,n,N)$, $n=N/2-S^z$ 
of the double-tetrahedra spin chain with a Hamiltonian given by the sum of
the terms in Eqs.~(\ref{2.01})
and (\ref{5.10})
the following dominant contribution at low temperatures for the magnetization $S^z=N/2-n_{\max}=N/4$
\begin{eqnarray}
\label{5.12}
Z_{{\rm{lm}}}(T,n_{\max},N)
=e^{-\frac{E_{\rm{FM}}-{\cal{N}}\varepsilon}{T}}
\prod_{\kappa}2\cosh\frac{\epsilon^{(2)}\cos\kappa}{T},
\end{eqnarray}
$\kappa=2\pi l/{\cal{N}}$, 
$l=-{\cal{N}}/2, -{\cal{N}}/2+1,\ldots,{\cal{N}}/2-1$
(we assume without loss of generality that ${\cal{N}}$ is even).
Low-temperature thermodynamic quantities in the thermodynamic limit ${\cal{N}}\to\infty$ follow from the free energy 
\begin{eqnarray}
\label{5.13}
\frac{F_{{\rm{lm}}}\left(T,n_{\max},N\right)}{\cal{N}}
=
\frac{E_{\rm{FM}}}{{\cal{N}}}-\varepsilon
\nonumber\\
-\frac{T}{2\pi}\int_{-\pi}^{\pi} d\kappa \ln\left(2\cosh\frac{\epsilon^{(2)}\cos\kappa}{T}\right).
\end{eqnarray}

Note that  the perturbation (\ref{5.02}) can also be included.
Then the formula (\ref{5.13}) has to be slightly modified,
namely, 
$2\epsilon^{(2)}\cos\kappa$ in Eq.~(\ref{5.13}) has to be replaced by $\epsilon^{(1)}+2\epsilon^{(2)}\cos\kappa$.
Thus, including both perturbations (\ref{5.02}) and (\ref{5.10})
we arrive at an effective (pseudo)spin-1/2 $XX$ chain in a transverse field
which governs the low-temperature physics of the double-tetrahedra chain in the subspace with $n=n_{\max}={\cal{N}}$ magnons
(i.e., at the magnetization $S^z=N/2-n_{\max}=N/4$).
It might be interesting to mention here
that the spin-1/2 $XX$ chain in a transverse field 
also emerges as an effective low-energy model for the diamond spin chain at high magnetic fields
if the conditions ensuring the presence of localized magnons become slightly violated.\cite{effective_xy}
For such a generalized diamond chain 
the spin-1/2 transverse $XX$ chain describes a weak spreading of the independent localized magnons over the whole chain.
Another related model, 
the so-called spin-chirality model, 
is used for effective description of three-leg spin tubes 
within the perturbation theory approach from the strong rung-coupling limit,
see Ref.~\onlinecite{sakai} and references therein.
We stress here that in the case at hand
the spin-1/2 transverse $XX$ chain describes propagating of chirality over the whole chain 
in the localized-magnon crystal state.

We consider next the frustrated triangular tube again in the subspace with $n=n_{\max}={\cal{N}}/2$ magnons 
(i.e., at the $\langle M\rangle=2/3$ plateau).
The ground state, 
besides the degeneracy owing to chirality, 
is two-fold degenerate, 
i.e., the independent localized magnons may occupy either even-site or odd-site sublattice only.
The perturbation to lift the degeneracy of the localized-magnon crystal state with respect to chirality 
that has to be added to the Hamiltonian (\ref{2.01}) corresponds to
${\cal{H}}_{\rm{s}}^{(2)}=\epsilon^{(2)}\sum_{m=1}^{{\cal{N}}}\left(\tau_m^{+}\tau_{m+2}^{-}+{\rm{H.c.}}\right)$,
i.e., it represents  now an $XX$ interaction between next-nearest-neighbor (pseudo)spins. 
In the initial spin model 
it is a certain four-site interaction which contains the sites of next-nearest-neighbor cells.

Finally we discuss briefly corresponding perturbations for the electron systems 
in the sector of $n={\cal{N}}$ electrons.\cite{footnote-hubb}
To have exactly one electron per cell
we introduce an extra repulsion between electrons in neighboring cells 
(i.e., we consider an extended Hubbard model).
Since the chirality and spin of electron states in each cell are not fixed, 
we have a $4^{\cal{N}}$-fold degenerate ground state.
A perturbation that lifts the degeneracy owing to chirality 
(independently of the spin) 
again corresponds to an $XX$ interaction between (pseudo)spins given by
${\cal{H}}_{\rm{e}}^{(2)}=\sum_{\sigma=\uparrow,\downarrow} {\cal{H}}_{0,\sigma}^{(2)}$,
${\cal{H}}_{0,\sigma}^{(2)}=\epsilon^{(2)}\sum_{m=1}^{{\cal{N}}}\left(\tau_m^{+}\tau_{m+1}^{-}+{\rm{H.c.}}\right)$,
where $\tau_m^{\pm}$ are defined by Eqs.~(\ref{5.09}) and (\ref{2.08}). 
Note that there are no spin indeces in the r.h.s. of the formula for ${\cal{H}}_{0,\sigma}^{(2)}$,
i.e., each state of the perturbed Hamiltonian is $2^{\cal{N}}$-fold degenerate owing to the electron spin.
Thus, 
if the perturbation interaction $H_{\rm{e}}^{(2)}$
(which contains, generally speaking, four-site terms in the electron-electron interaction between the neighboring cells)
is switched on,
the thermodynamic properties of the extended Hubbard model with $n={\cal{N}}$ electrons on both considered lattices 
are related to those of the (pseudo)spin-1/2 $XX$ chain, 
see the corresponding results for the spin model, Eqs.~(\ref{5.12}) and (\ref{5.13}).

\section{Conclusions}
\label{sec6}
\setcounter{equation}{0}

To summarize,
we have considered the low-temperature properties of the spin-1/2 antiferromagnetic Heisenberg model and the repulsive Hubbard model 
on two 1D lattices containing equilateral triangles.
The lattices under consideration have a dispersionless lowest-energy band for the one-particle problem,
and the corresponding localized one-particle states can be trapped on the triangles.
Due to the triangular geometry of the trapping cells 
the localized one-particle states are characterized by two possible values of the chirality.
Using the localized nature of the one-particle states
we can construct corresponding many-particle low-energy states.
Moreover, we estimate their contribution to thermodynamics 
exploiting  classical lattice-gas description of the low-energy degrees of freedom of the quantum models. 
The lattice-gas description yields explicit analytical formulas for thermodynamic quantities 
at low temperatures in a certain region of the magnetic field (chemical potential) 
for the spin (electron) model.   
We investigate  the  effects of the localized states on the low-temperature thermodynamics.
In detail we discuss the specific heat $C(T,h,N)$ for the spin systems 
and 
the entropy $S(T,n,N)=\int_0^{T}dT^{\prime}C(T^{\prime},n,N)/T^{\prime}$ for the electron systems.
Both quantities exhibit fingerprints of highly-degenerate localized states,
namely, 
additional low-temperature peaks of $C(T,h,N)/N$
or 
a finite residual ground-state entropy $S(T=0,n,N)/N$.
Since the considered systems show a significant zero-temperature entropy,
they may exhibit an enhanced magnetocaloric effect.\cite{mce}

The degeneracy related to the chirality degrees of freedom may be lifted by small symmetry-breaking interactions.
For the perturbed system we provide an effective description 
of low-energy degrees of freedom of the considered spin and electron models 
in terms of a (pseudo)spin-1/2 $XX$ chain in a transverse field.
It might be interesting to note
that in contrast to usual cases, 
where the spins are related to the spin degree of freedom of electrons,
the (pseudo)spins emerging in our case are related to the charge degree of freedom of electrons 
and they simply stand for a (pseudo)spin representation of the chirality.
Moreover,
the chirality inherent in the considered spin models on geometrically frustrated lattices 
may also give rise to a chain of quantum (pseudo)spins 1/2. 

It is worthy noting 
that quantum spin chains are often used in quantum information theory 
both for illustration of basic concepts and as candidates for physical implementation.\cite{quant_inf}
From such a perspective, 
manipulation with chirality\cite{chirality} realized in (pseudo)spin chains 
may be an interesting subject for further studies.
Finally,
although the main advantage of the considered strongly correlated lattice models 
is the possibility to elaborate a theoretical description of thermodynamics
which works perfectly well at low temperatures for high magnetic fields or low concentrations of electrons,
we may mention here some experimental solid-state realizations of similar systems,
see Refs.~\onlinecite{yamamoto,nedko-cmp,nedko}.

\section*{Acknowledgments}

The numerical calculations were performed using 
ALPS package\cite{alps}
and
J.~Schulenburg's {\it spinpack}.\cite{spinpack}
The authors thank A.~Honecker for fruitful discussions.
The present study was supported by the DFG (projects Ri615/18-1 and Ri615/19-1).
M.~M. acknowledges the kind hospitality of the University of Magdeburg in 2010.
O.~D. acknowledges the kind hospitality of the University of Magdeburg in 2010 and 2011.


\begin{thebibliography}{999}
\bibitem{footnote1}
Each site of the auxiliary lattice may be either empty or occupied by the localized magnon 
with two possible values of the chirality.
\bibitem{footnote2}
To find 
$\sum_{n=0}^{{\cal{N}}/2}g_{{\cal{N}}}(n)=\sum_{n=0}^{{\cal{N}}/2}2^n{\cal{Z}}_{{\rm{hd}}}(n,{\cal{N}})$
for ${\cal{N}}\to\infty$
we note that the required number is given by
$\sum_{n=0}^{{\cal{N}}/2}z^n{\cal{Z}}_{{\rm{hd}}}(n,{\cal{N}})
\underset{{\cal{N}}\to\infty}{\longrightarrow}
[(1+\sqrt{1+4z})/2]^{\cal{N}}$
at $z=2$.
\bibitem{footnote-hubb}
Note
that the subspace of $n=n_{\max}=2{\cal{N}}$ electrons is not of interest in the context of the considered issue,
since in the ground state all ${\cal{N}}$ cells are occupied by two electrons with different chiralities.
The degeneracy of this ground state is $3^{\cal{N}}$ 
and is connected with three components of the triplet state at each cell.
\end{thebibliography}
\end{document}